\documentstyle[sprocl,epsfig]{article}

\def\Journal#1#2#3#4{{#1} {\bf #2}, #3 (#4)}

\def\NPB{{\em Nucl. Phys.} B}
\def\PLB{{\em Phys. Lett.}  B}
\def\PRL{\em Phys. Rev. Lett.}
\def\PRD{{\em Phys. Rev.} D}

\def \be  {\begin{equation}}
\def \ee  {\end{equation}}
\def \ba  {\begin{eqnarray}}
\def \ea  {\end{eqnarray}}
\def \baa {\begin{eqnarray*}}
\def \eaa {\end{eqnarray*}}
\def \bb  {\begin{thebibliography}}
\def \eb  {\end{thebibliography}}

\def \lab #1 {\label{#1}}

\def \fracs #1#2 {\mbox{\small $\frac{#1}{#2}$}}

\def \bin #1#2 {{\left({#1}\atop{#2}\right)}}
\def \as {\relax\ifmmode\alpha_s\else{$\alpha_s${ }}\fi}

\def \al #1 {\frac {\as({#1})}{\pi} }
\def \ds #1 {\ooalign{$\hfil/\hfil$\crcr$#1$}}

\begin{document}
%
\thispagestyle{empty}
\onecolumn
\begin{flushright}
\large
NIKHEF/00-14 \\
YITP-00-33 \\
RIKEN-BNL preprint \\
June 2000
\end{flushright}
\vspace{1.2cm}
 
\renewcommand{\thefootnote}{\fnsymbol{footnote}}
\setcounter{footnote}{1}
\begin{center}
\begin{LARGE}
\begin{bf}
Current Issues in \\ Prompt Photon Production\footnote{
Invited talk presented by W.~Vogelsang 
at the ``8th International Workshop on Deep-Inelastic 
Scattering'' (DIS2000), 25-30 April 2000, Liverpool, UK\\}
 
\end{bf}
\end{LARGE}

\vspace*{1.2cm}
{\Large  Eric~Laenen$^a$, George~Sterman$^b$, Werner~Vogelsang$^{c,}$\footnote{
WV is grateful to RIKEN, Brookhaven National Laboratory and the U.S.
Department of Energy (contract number DE-AC02-98CH10886) for
providing the facilities essential for the completion of this work}}
 
\vspace*{4mm}
 
\begin{large}
$^a$ NIKHEF Theory Group, Kruislaan 409\\ 1098 SJ Amsterdam, The
Netherlands 
\vskip 3mm

$^b$ C.N.\ Yang Institute for Theoretical Physics,
SUNY Stony Brook\\
Stony Brook, New York 11794 -- 3840, U.S.A. 
\vskip 3mm

$^c$ {RIKEN-BNL Research Center, Brookhaven National Laboratory}
\vspace*{2mm}
Upton, NY 11973, U.S.A.\\[3pt]

\end{large}
\vspace*{1.4cm}
 
%
 
{\Large \bf Abstract}
\end{center}
\vskip 3mm
 
\noindent
We give a brief account of recent theoretical 
developments in prompt photon production.
 
\normalsize
\newpage        

\title{CURRENT ISSUES IN PROMPT PHOTON PRODUCTION}
\vskip -1mm
\author{E. LAENEN}

\address{NIKHEF Theory Group, Kruislaan 409, 1098 SJ Amsterdam, The
Netherlands}

\author{G. STERMAN}

\address{C.N.\ Yang Institute for Theoretical Physics,
SUNY Stony Brook\\ Stony Brook, New York 11794 -- 3840, U.S.A.}  

\author{W. VOGELSANG}

\address{RIKEN-BNL Research Center, Brookhaven National Laboratory,\\
Upton, NY 11973, U.S.A.}

\maketitle\abstracts{We give a brief account of recent theoretical 
developments in prompt photon production.}
\vskip -2mm
Prompt-photon production at high transverse momentum~\cite{photondata}, 
$pp,p\bar{p},pN\rightarrow \gamma X$, has been a classical tool for 
constraining the nucleon's gluon density, since at leading order a photon 
can be produced in the Compton reaction $qg\to\gamma q$. The `point-like' 
coupling of the photon to the quark provides a supposedly `clean' 
electromagnetic probe of QCD hard scattering. In the framework of QCD 
perturbation theory, the inclusive cross section is written in a factorized 
form:
\be
\frac{d \sigma_{AB\to \gamma X}(x_T^2)}{dp_T } =
\sum_{ab} \phi_{a/A}(x_a,\mu)\, \otimes \phi_{b/B}(x_b,\mu)\; \otimes
\frac{d\hat \sigma_{ab\to \gamma c}\left(\hat
x_T^2,\mu\right)}{dp_T}\; .
\label{dgamptcofact}
\ee
For simplicity, we have integrated over the rapidity $\eta$ of the prompt 
photon. In~(\ref{dgamptcofact}), $\phi_{a/A}(x_a,\mu)$ denotes the parton 
density for species $a$ in hadron $A$, at factorization scale $\mu$, 
$x_a$ being the parton's momentum fraction. $d\hat\sigma_{ab\to\gamma c}
/dp_T$ are the partonic hard-scattering functions, which have been calculated 
to next-to-leading order~\cite{dgnlo}. Hadronic and partonic scaling 
variables are $x_T^2 \equiv { 4p_T^2/S}$ and $\hat x_T^2 \equiv 
{ 4p_T^2/\hat s}$, respectively, with $\hat s=x_ax_bS$.

Unfortunately, in experiment, one has to deal with a substantial
background of photons from $\pi^0$ decay. In addition, high-$p_T$ 
photons can be produced in jets, when a parton, resulting 
from a pure QCD reaction, fragments into a photon plus a number of hadrons. 
This inevitably introduces dependence on non-perturbative (parton-to-photon) 
fragmentation functions. So far, the latter are insufficiently known. 
Theoretical studies~\cite{aurenche93,gluck94,vogt95,aurenche99},
based on predictions~\cite{aurenche93,gluck93,bourhis98} for the photon 
fragmentation functions, indicate that the fragmentation component 
is in practice a subdominant, albeit non-negligible, effect. In the 
fixed-target regime, fragmentation photons are believed~\cite{vogt95} to 
contribute at most $20\%$ to the direct photon cross section. At collider 
energies, the fragmentation mechanism can easily produce about half or more 
of the observed photons~\cite{aurenche90,aurenche93,vogt95}.
Here an `isolation' cut is imposed on the photon signal in experiment, 
in order to suppress the $\pi^0$ background. Isolation is usually
realized by drawing a cone of fixed aperture in $\varphi$--$\eta$ space 
around the photon, restricting the hadronic transverse energy 
allowed in this cone to a certain small fraction $\varepsilon$ 
of the photon transverse energy. In this way, the fragmentation 
contribution, resulting from an essentially collinear process, is
diminished~\cite{berger91}, probably to a level of 15--20\%, or 
less~\cite{gluck94,vogt95}.

We mention that subtleties were observed~\cite{bgq,aur97,catani} in the 
past concerning the factorizability of the {\em isolated} prompt photon 
cross section. While in~\cite{catani} factorization was eventually proved 
to hold, it was also shown~\cite{catani,aur97,bgq} that with isolation 
the NLO partonic hard scattering function has a Sudakov-type singular behavior 
at a certain point {\em inside} the physical region, requiring soft-gluon 
resummations in order to obtain a reliable theoretical result. It was also 
pointed out~\cite{berger91} that the (in practice~\cite{cdf}) small size 
of $\varepsilon$ can give rise to potentially large logarithms of 
$\varepsilon$, associated with soft-gluon emission into the isolation cone. 
Further work on the fragmentation component is clearly needed. A recent 
suggestion~\cite{frixione98} to refine isolation by allowing less and less 
hadronic energy the closer to the photon it is deposited, will, at least 
theoretically, eliminate the fragmentation component altogether and could 
potentially avoid some of the problems just mentioned. This isolation 
prescription has been applied in studies for prompt photon production at 
RHIC and LHC~\cite{fv}.

A pattern of disagreement between theoretical predictions and experimental 
data for prompt photon production has been observed in recent 
years~\cite{cdf,e706,ua6}, not globally curable by changing 
factorization and renormalization scales or by `fine-tuning' the gluon 
density~\cite{vogt95,huston95,aurenche99}. The most serious problems relate
to the fixed-target data, where NLO theory dramatically underpredicts 
some data sets~\cite{e706,ua6}. At collider energies, there is less reason 
for concern, but also here the agreement is not fully satisfactory. The mutual 
consistency of the various fixed-target data sets has been questioned 
in~\cite{aurenche99}. On the other hand, various improvements of the 
theoretical framework have been developed.

`Threshold' resummations for the inclusive prompt photon cross 
section have been performed in~\cite{LOS,CMN} and applied phenomenologically
in~\cite{CMNOV,KO}. As $\hat s$ approaches its minimum value at 
$\hat x_T^2=1$, corresponding to `partonic threshold' when the initial
partons have just enough energy to produce the high-$p_T$ photon and
the recoiling jet, the phase space available for gluon bremsstrahlung 
vanishes, resulting in corrections to $d\hat\sigma/dp_T$
as large as $\as^k \ln^{2k}(1-\hat x_T^2) \, \hat \sigma^{\rm Born}$ at
order $\as^k$ in perturbation theory. Threshold 
resummation~\cite{dyresum,LOS,CMN} organizes 
\begin{figure}[t]
\vskip -0.7cm
\begin{center}
\epsfig{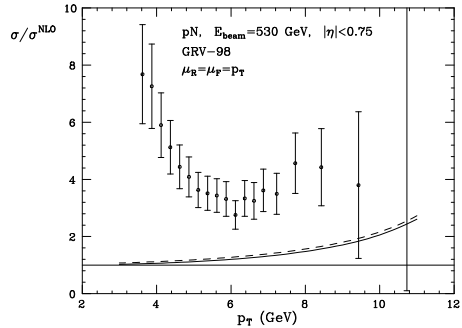}
\caption{Threshold-resummed prompt photon cross section based
on the formalisms of~$\!{}^{20,(25)}$ (solid) and~$\!{}^{21,(22)}$
(dashed), divided by the NLO cross section. Parton densities are 
from~$\!{}^{26}$. The E706 prompt photon data~$\!{}^{17}$ are also 
shown in the same normalization.\label{fig:thr}}
\end{center}
\vskip -6mm
\end{figure}
this singular, but integrable, 
behavior of $d\hat\sigma/dp_T$ to all orders in $\as$. It is carried out in 
Mellin-$N$ moment space, where the logarithms are of the form $\as^k \ln^{2k}
(N) \, \hat 
\sigma^{\rm Born} (N)$. Its application is particularly interesting 
in the fixed-target regime, since here the highest $x_T$ are attained in 
the data, and since here the discrepancy between data and theory is 
largest. As seen from Fig.~\ref{fig:thr}, one obtains a significant, albeit 
not sufficient, enhancement of the theory prediction. Also, a dramatic 
reduction of scale dependence is found~\cite{CMNOV}. One notices that the 
formalisms of~\cite{LOS,CMN}, which differ in so far as the one of~\cite{LOS} 
resums at fixed photon rapidity, whereas the one of~\cite{CMN} applies 
to the cross section fully integrated over rapidity, yield very similar 
predictions in practice.

Valuable insight was gained in studies of direct photon production in 
which transverse smearing of the momenta of the initial partons 
was incorporated~\cite{huston95,apan96,martin98,LiLai,kimber99}.
If, say, in the Compton process $qg\to \gamma q$ the initial partons have 
a non-zero $k_T$, the $\gamma\, q$ {\em pair} in the final state will 
acquire a net transverse momentum $Q_T$, which may make the process softer 
than it would be otherwise and result in an enhancement in the photon 
$p_T$ spectrum. The first approaches~\cite{huston95,apan96,martin98} assumed 
Gaussian dependence on $k_T$ and enjoyed phenomenological success in that 
they were able to bridge the large gaps between data and theory 
for appropriate choices of average $\langle k_T \rangle$, guided by 
values of dimuon, dijet and diphoton {\em pair} transverse momenta measured 
in hadronic reactions. On the other hand, it was clear that a more developed 
theoretical framework was required. For instance, as acknowledged 
already in~\cite{apan96}, the physical origin for the parton $k_T$ 
dependence should be thought of as initial-state gluon emission, so that
$Q_T$ takes the role of pair recoil against unobserved radiation. 
Ideally, the phenomenologically required 
$k_T$ smearing would have to be understood in 
terms of a $Q_T$-resummation calculation, with perturbative as 
well as non-perturbative components, as familiar from the well-explored 
case of Drell-Yan dimuon production, where soft-gluon radiation gives rise 
to powers of large logarithms at small $Q_T$, which can 
be resummed to all orders. In the inclusive partonic prompt photon cross 
section $d \hat \sigma_{ab\to \gamma c}/dp_T$, such $Q_T$-logarithms will 
not be visible at any given order of perturbation theory; however, they will 
show up when considering $d \hat \sigma_{ab\to \gamma c}/ d^2 Q_T\, dp_T$ 
and can be resummed prior to $Q_T$ integration. Attempts in this direction 
were first made in~\cite{LiLai,kimber99}, where parton densities, unintegrated 
over parton transverse momentum, but constructed from the usual 
densities, were used. The `double-logarithmic approximation' was employed 
in~\cite{LiLai,kimber99}. In~\cite{kimber99}, only small recoil effects
were found when a strong ordering constraint was imposed.

In~\cite{LSV}, a simultaneous resummation in both threshold and transverse 
momentum logarithms was achieved. The possibility of doing this had previously
been pointed out in~\cite{Liunified}. Contributions to the hard 
scattering function associated with threshold resummation are redistributed 
over soft gluon transverse momenta, simultaneously conserving energy 
{\em and} transverse momentum. Large logarithmic corrections in $Q_T$ to 
$d\hat \sigma_{ab\to \gamma c}/ d^2 Q_T\, dp_T$ arising in the threshold 
region are resummed jointly with threshold logarithms. A further crucial and 
distinct feature of~\cite{LSV} is that it remains within the formalism of 
collinear factorization, which implies use of ordinary parton densities. 

The possibility of joint resummation for singular behavior in $Q_T$ and 
$1-\hat x_T^2$ is ensured by the factorization properties of the partonic 
cross section near threshold~\cite{cttwcls}. Singular $Q_T$ behavior 
is organized in impact parameter $b$-space, where the logarithms exponentiate.
Defining `profile functions' in $Q_T$ as 
\vskip -1mm
\be
P_{ij}\left( N,{\bf Q}_T,Q,\mu \right)=\int d^2 {\bf b} \,
{\rm e}^{-i {\bf b} \cdot {\bf Q}_T} \,
\exp\left[E_{ij\to \gamma k}\left( N,b,Q,\mu \right)\right]\, ,
\label{Pdef}
\ee
where the resummed exponents $E_{ij\to \gamma k}$ can be found in~\cite{LSV}
to NLL in both $b$ and $N$, the resulting resummed cross section is given in 
terms of Mellin moments of the parton distributions and of the squared 
Born amplitudes $|M_{ij}|^2$ as
\ba
{p_T^3 d \sigma^{({\rm resum})}_{AB\to \gamma X} \over dp_T}
&=& \sum_{ij} \frac{p_T^4}{8 \pi S^2} \int_{\cal C} {dN \over 2 \pi i}\;
\tilde{\phi}_{i/A}(N,\mu) \tilde{\phi}_{j/B}(N,\mu)\;
\int_0^1 d\tilde x^2_T \left(\tilde x^2_T \right)^N \nonumber \\
&&\hspace{-25mm} \times
{|M_{ij}(\tilde x^2_T)|^2\over \sqrt{1-\tilde{x}_T^2}}
\int {d^2 {\bf Q}_T \over (2\pi)^2}\;
\Theta\left(\bar{\mu}-Q_T\right)
\left( \frac{S}{4 {\bf p}_T'{}^2} \right)^{N+1}\;
P_{ij}\left( N,{\bf Q}_T,\frac{2 p_T}{\tilde x_T},\mu \right)\, ,
\label{1pIresumE}
\ea
the recoil effect entering through ${\bf p}_T'={\bf p}_T-
{\bf Q}_T/2$. Eq.\ (\ref{1pIresumE}) reverts to the threshold-resummed
cross section for ${\bf p}_T'\to{\bf p}_T$. $Q_T$ is limited by the scale 
$\bar{\mu}$, 
to avoid going outside the region where the singularities in 
$Q_T$ dominate.  

\begin{figure}[t]
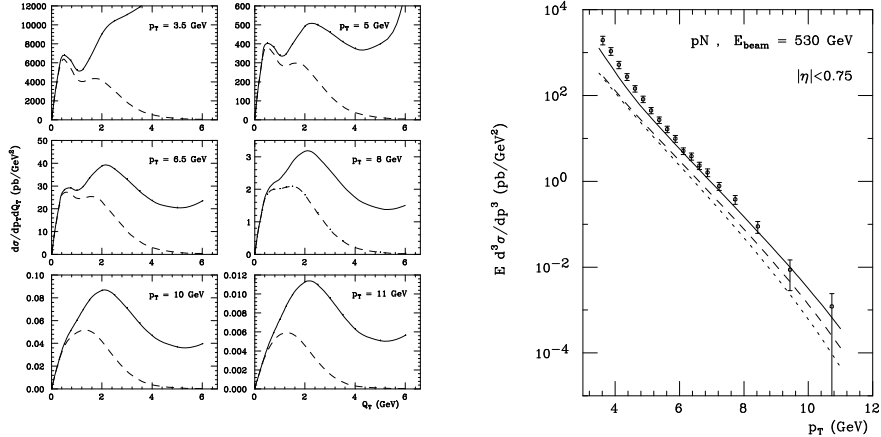

\vskip -0.7cm
\epsfig{figure=qtshape.eps,height=5.5cm,angle=90}
\vskip -5.5cm
\hskip 6.4cm
\epsfig{figure=e706crs.eps,height=5.2cm,angle=90}
\vskip -0.2cm
\caption{Left: prompt photon cross section
$d\sigma_{{\rm pN}\to\gamma X}/dQ_Tdp_T$ at {\protect $\sqrt{s}=31.5$} GeV.
Dashed lines are computed without recoil,
solid lines are with recoil. Right: $Ed^3 \sigma_{{\rm pN}\to\gamma X}/dp^3$. 
The dotted line represents the full NLO calculation, while the dashed and 
solid lines respectively incorporate pure threshold resummation 
(see~$\!{}^{21}$ and Fig.~\ref{fig:thr}) and the joint resummation 
of~$\!{}^{31}$.
\label{fig:f2}}
\vskip -6mm
\end{figure}

Cross sections computed on the basis of~(\ref{1pIresumE}) are shown in 
Fig.\ \ref{fig:f2} as functions of $Q_T$ at fixed $p_T$.
The kinematics are those of the E706 experiment~\cite{e706}; see~\cite{LSV}
for details of the calculation, in particular those regarding the evaluation
of the $b$-integral in~(\ref{Pdef}). The  dashed lines are 
$d\sigma^{(\rm resum)}_{{\rm pN}\to\gamma X}/dQ_Tdp_T$, with recoil 
neglected by fixing ${\bf p}_T'={\bf p}_T$, thus showing how each $Q_T$ 
contributes to threshold enhancement. The solid lines show the same, but 
now including the true recoil factor $(S/4{\bf p}_T'{}^2)^{N+1}$.
The resulting enhancement is clearly substantial. For small $p_T$, the
enhancement simply grows with $Q_T$, while for $p_T$ above 5 GeV it has a 
dip at about $Q_T = 5$ GeV, which remains substantially above zero. This 
makes it difficult to confidently determine $\bar\mu$.

So far, the numerical results given in~\cite{LSV} are primarily to be 
regarded as illustrations, rather than quantitative predictions. This applies
in particular to the resummed $Q_T$-{\em integrated} cross section, also 
shown in Fig.~\ref{fig:f2} for $p_T\ge 3.5$ GeV and $\bar\mu=5$ GeV. 
Comparison with Fig.~\ref{fig:thr} demonstrates the size of the additional
enhancement that recoil can produce and its potential phenomenological 
impact. Further work on the implementation of practical nonperturbative 
estimates and of matching procedures is required.
%
%

\end{document}